\documentclass[sigchi-a]{acmart}
\usepackage{booktabs} 
\usepackage{ccicons}  

\setcopyright{none}

\begin{document}
\title{Lab Hackathons to Overcome Laboratory Equipment Shortages in Africa: Opportunities and Challenges}

\author{Helena Webb}
\affiliation{%
  \institution{University of Oxford}
  \city{Oxford}
  \postcode{OX1 3QD}
  \country{UK}}
\email{helena.webb@cs.ox.ac.uk}

\author{Jason R.C. Nurse}
\affiliation{%
  \institution{University of Kent}
  \city{Canterbury}
  \postcode{CT2 7NZ}
  \country{UK} }
\email{j.r.c.nurse@kent.ac.uk}

\author{Louise Bezuidenhout}
\affiliation{%
  \institution{University of Oxford}
  \city{Oxford} \postcode{OX1 2JD} \country{UK}}
\email{louise.bezuidenhout@insis.ox.ac.uk}

\author{Marina Jirotka}
\affiliation{\institution{University of Oxford}
  \city{Oxford} 
  \postcode{OX1 2JD} \country{UK}}
\email{marina.jirotka@cs.ox.ac.uk}

\copyrightyear{} 
\acmYear{} 
\acmConference{}
\acmBooktitle{}
\acmPrice{} 
\acmDOI{}
\acmISBN{}

%
%


\begin{abstract}
Equipment shortages in Africa undermine Science, Technology, Engineering and Mathematics (STEM) Education. We have pioneered the LabHackathon (LabHack): a novel initiative that adapts the conventional hackathon and draws on insights from the Open Hardware movement and Responsible Research and Innovation (RRI). LabHacks are fun, educational events that challenge student participants to build frugal and reproducible pieces of laboratory equipment. Completed designs are then made available to others. LabHacks can therefore facilitate the open and sustainable design of laboratory equipment, in situ, in Africa. In this case study we describe the LabHackathon model, discuss its application in a pilot event held in Zimbabwe and outline the opportunities and challenges it presents.
\end{abstract}

\newpage

\maketitle

\begin{marginfigure}
    \includegraphics[width=\marginparwidth]{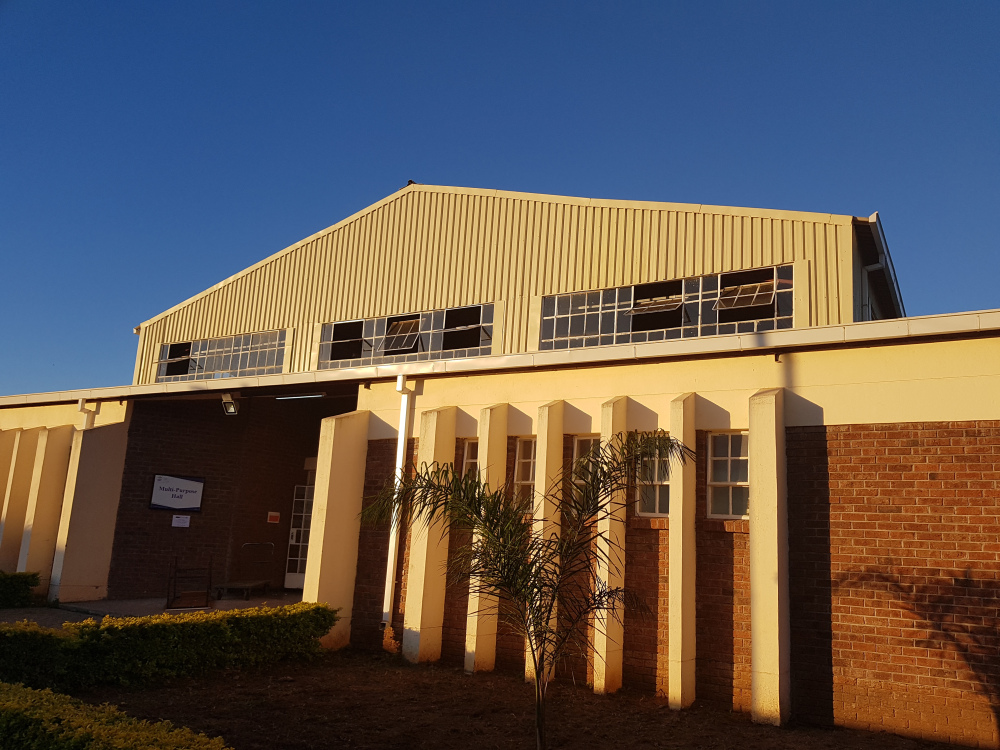}
   \caption{Harare Institute of Technology: the venue for the pilot LabHack event in Zimbabwe}
    \label{fig:venue}
    \vspace{2em}    
\end{marginfigure}

\begin{marginfigure}
    \includegraphics[width=\marginparwidth]{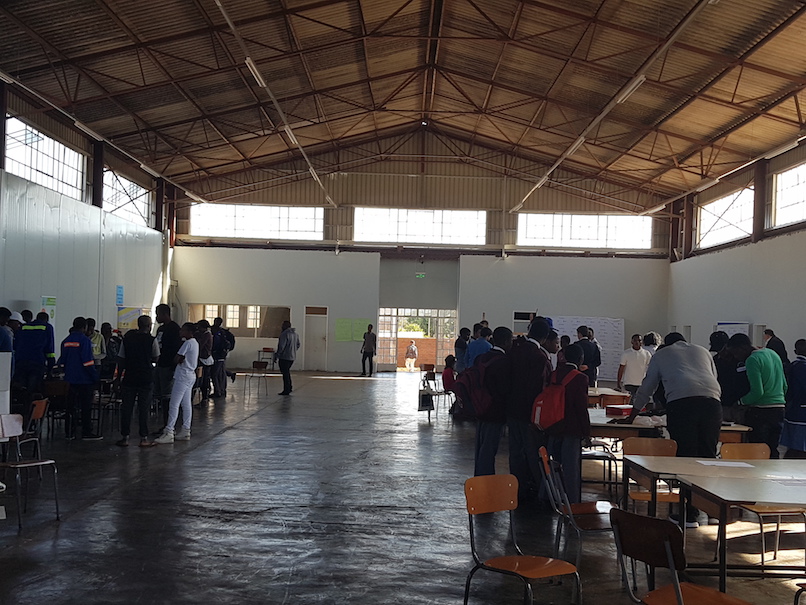}
    \caption{Inside the hall used for equipment demonstrations during the pilot LabHack event}
    \label{fig:venue2}
\end{marginfigure}

\section{Introduction}
Science, Technology, Engineering and Mathematics (STEM) education in Africa is challenged by shortages of laboratory equipment in teaching institutions.  Large numbers of students often have to share single items of equipment, making it difficult for them to gain the hands-on experience necessary for effective STEM education.  Moreover, the dominance of laboratory equipment designed by and for the Global North~\cite{rgsnd} means that equipment is often difficult to use, maintain and repair within an African context.  Thus, many African learning laboratories remain critically under-resourced and reliant on the efforts of key champions to access inter/national funds for equipment purchase~\cite{bezuidenhout2017100}.  

The emerging field of Open Hardware~\cite{openhardware} offers alternatives to reliance on expensive, proprietary laboratory equipment.  The Open Hardware movement includes a global drive towards the redesign and free dissemination of equipment plans and has already produced a wide range of online resources relating to the production of laboratory equipment in situ from scratch. At the same time, Responsible Research and Innovation (RRI) initiatives  have successfully demonstrated the value of conducting research and innovation with, and for, society~\cite{von2013vision}. RRI approaches typically involve a range of stakeholders in activities that align the outcomes of innovation with the values of society. They can therefore provide creative, long-term solutions to seemingly intractable problems.

\begin{marginfigure}
	\vspace{2em}    
	\includegraphics[width=\marginparwidth]{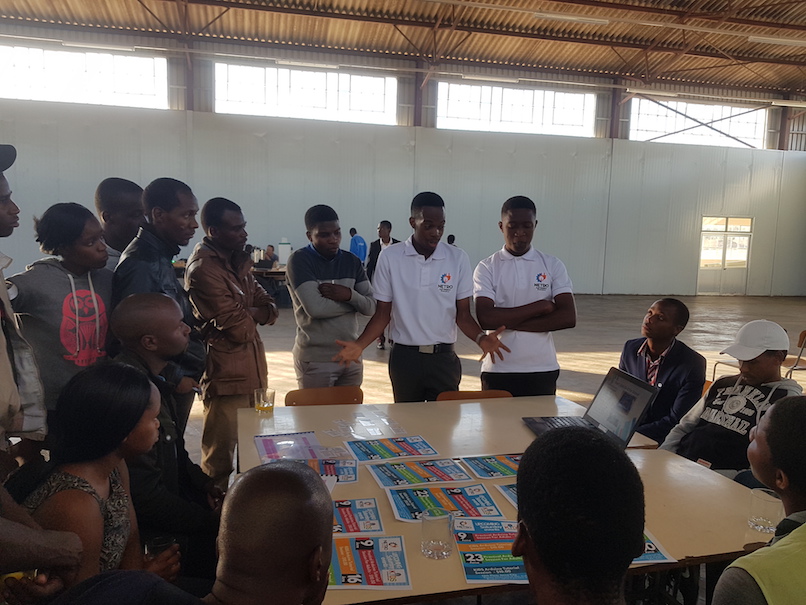}
	\caption{At the LabHack, participants join a workshop session on using an Arduino, which was run by a local start-up company}
	\label{fig:workshop}
	\vspace{2em}   
\end{marginfigure}

\begin{marginfigure}
	\includegraphics[width=\marginparwidth]{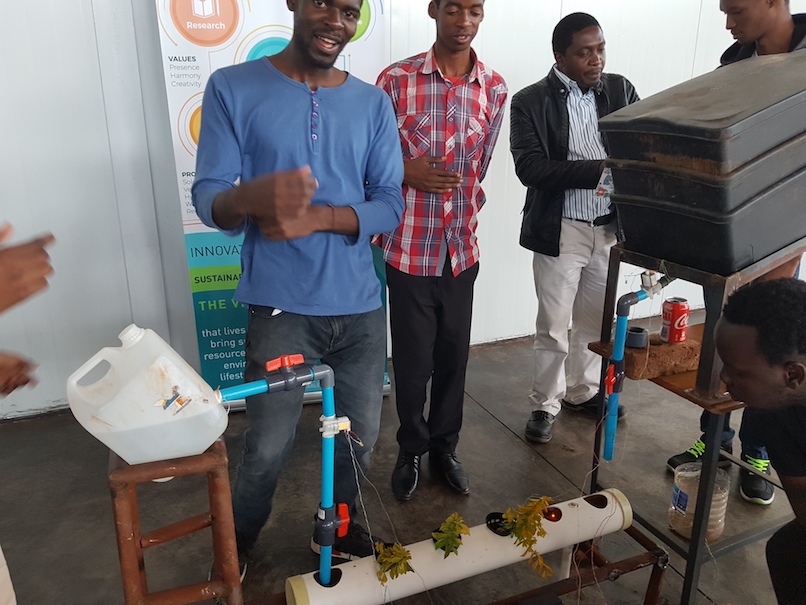}
	\caption{A team presenting their prototype hydroponics and vermiculture infusion instrument at the LabHack}
	\label{fig:Irre}
\end{marginfigure}

\begin{marginfigure}
	\includegraphics[width=\marginparwidth]{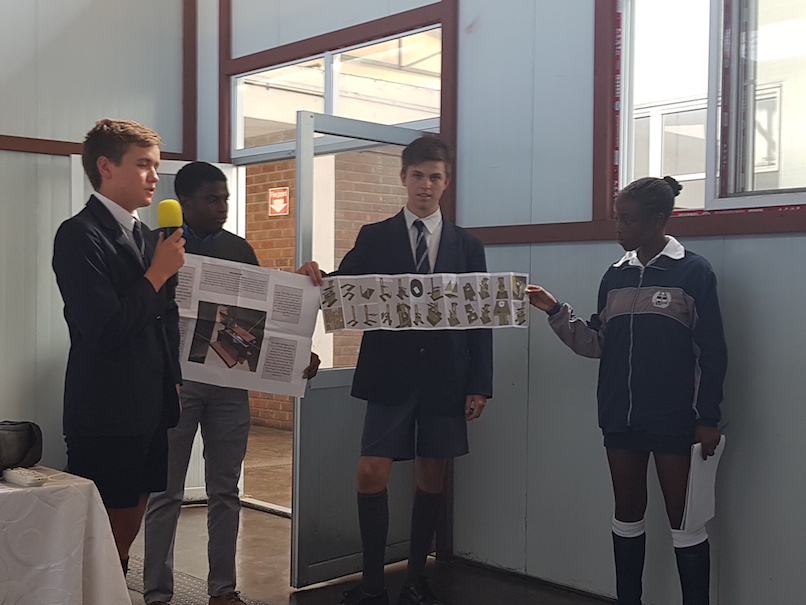}
	\caption{Participants present and describe their work}
	\label{fig:presentationwork}
\end{marginfigure}

\begin{marginfigure}
	\vspace{2em}
	\includegraphics[width=\marginparwidth]{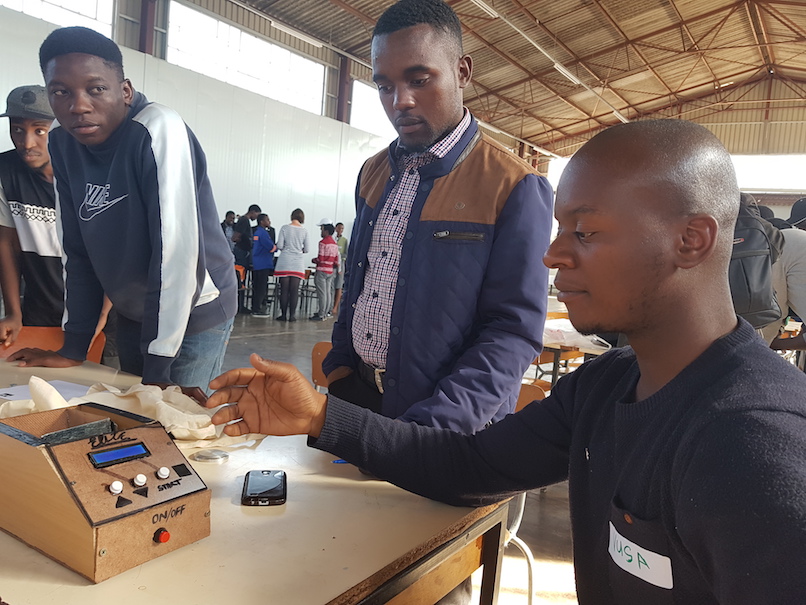}
	\caption{A participant demonstrates his team's PCR machine. This team won the prize for best documentation}
	\label{fig:pcrm}
\end{marginfigure}

\begin{marginfigure}
	\vspace{2em}
	\includegraphics[width=\marginparwidth]{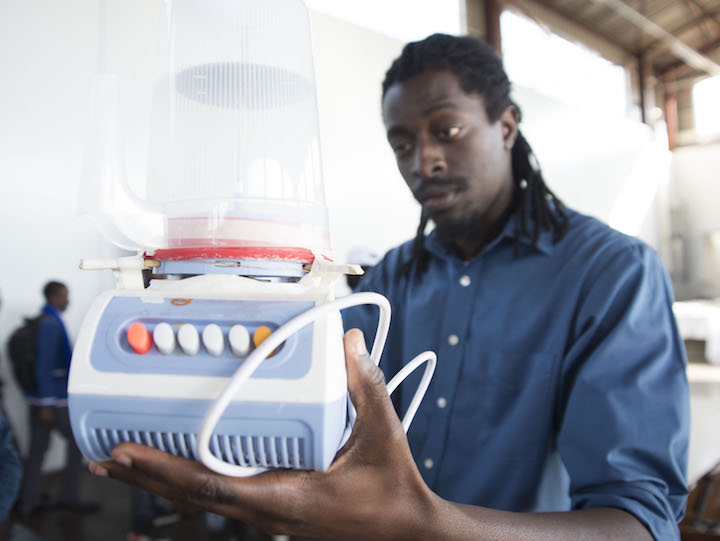}
	\caption{The team that won the prize for most frugal design built a centrifuge from a food mixer. Photo credit to Jeffrey Barbee/Alliance Earth}
	\label{fig:frugal}
\end{marginfigure}

\begin{marginfigure}
	\vspace{1em}
	\includegraphics[width=\marginparwidth]{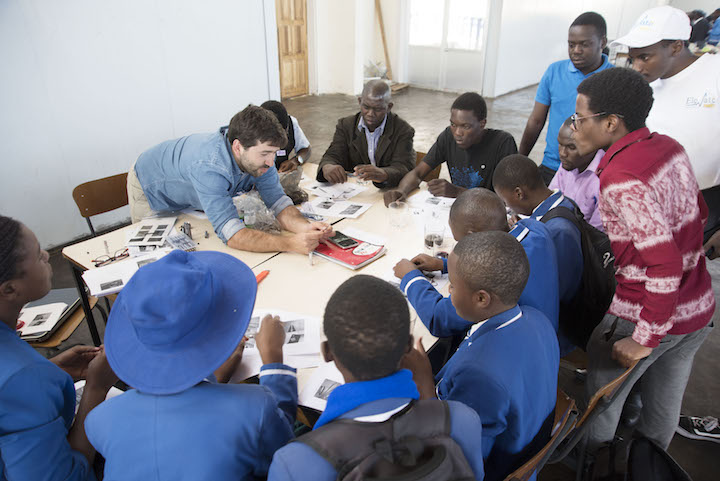}
	\caption{Participants take part in a session on building a \$10 microscope. The session was run by event collaborator Andre Maia Chagas. Photo credit to Jeffrey Barbee/Alliance Earth}
	\label{fig:microscope}
\end{marginfigure}

\begin{marginfigure}
	\vspace{2em}    
	\includegraphics[width=\marginparwidth]{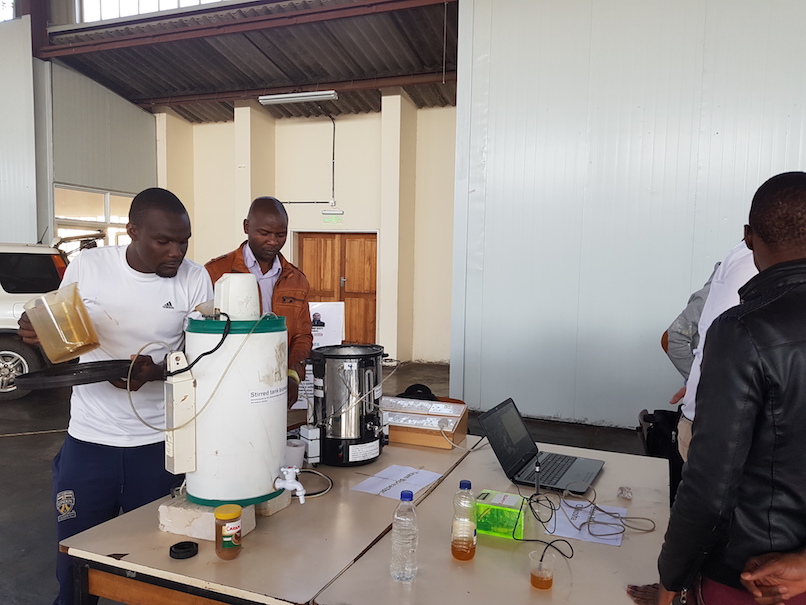}
	\caption{A team presenting their prototype bioreactor}
	\label{fig:Bioreactor}
	\vspace{2em}   
\end{marginfigure}

This case study describes the LabHackathon (or LabHack), a laboratory equipment hackathon that we have developed and trialled as an approach to address laboratory equipment shortages in Africa. Following a brief background section, our methods section describes the LabHack model and our findings section analyses and considers the pilot event held in Zimbabwe in June 2018 (photos of the event can  be seen in Figures~\ref{fig:venue}-\ref{fig:group2}). Next, our discussion section outlines the lessons we have learned from the pilot and other implications for practice. The LabHack model combines innovations in the traditional hackathon experience~\cite{porter2017reappropriating} with insights from the Open Hardware and RRI movements. It motivates and challenges students to design and build their own frugal, reproducible pieces of laboratory equipment. 

They do so by drawing on available Open Hardware resources and then make their own design plans available to others.  In this way, LabHacks can be a novel mechanism to facilitate the open and sustainable design of key laboratory equipment in, for and by, Africans. 

\section{Background}
Hackathons are highly popular events that typically feature groups of individuals competing over a constrained period of time to achieve some task or solve a defined problem. While its origins have been linked to prototyping in Silicon Valley~\cite{lodato2016issue}, today hackathons are applied to several varied domains. A few examples include hackathons which target sustainability (or green hackathons, as they are called) \cite{zapico2013hacking} and  those facilitating the exploration of digital technologies for mental health and self-harm \cite{birbeck2017self}. These applications seek to bring the innovative, fluid, cooperative and marathon-like nature of hackathons to address and discuss issues across several domains. There have also been attempts to make hackathons multidisciplinary in order to draw on the expertise of participants from backgrounds such as law, marketing and business~\cite{porter2017reappropriating,taylor2018everybody}. Hackathons have become a valuable educational and practical tool across the CHI community but challenges exist around how to structure events to best meet the needs and interests of participants, and to achieve long term impact~\cite{pe20182}.

Within Africa, hackathons enable members of the public to engage with technology. There have been hackathons on social issues such as the developing sustainable bioinformatics capacity \cite{ahmed2018organizing}, as well as topics including cybersecurity \cite{cmufb18}. As hackathons have grown in popularity in Africa, so too has the Open Hardware movement. The Africa Open Science and Hardware Summit \cite{africaosh} ,which aims to create a community of individuals from across society interested in Open Science and Hardware, is a good example of this. There has also been research in this area examining the challenges researchers face and more broadly, the accessibility of scientific research in Africa \cite{bezuidenhout2017100}. Our work takes a novel approach in combining these two fields in order to address real-world sociotechnical issues.

\section{Design of LabHack model}

\begin{figure}
	\includegraphics[width=0.75\textwidth]{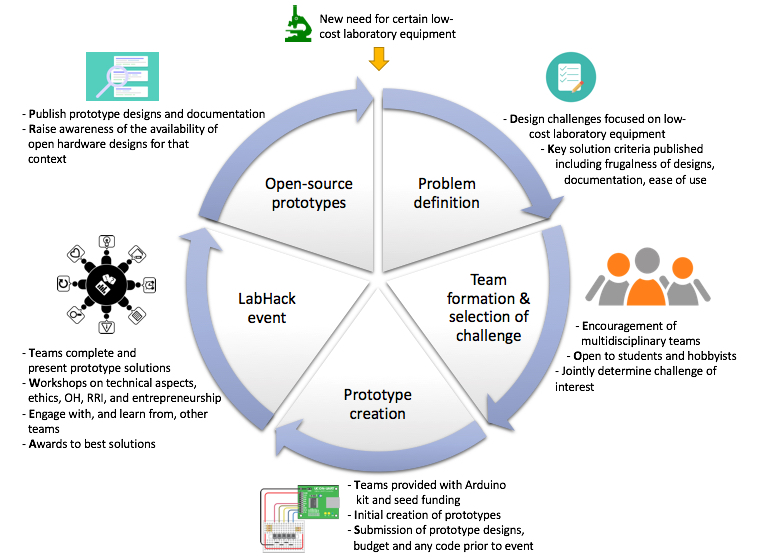}
	\caption{An overview of the LabHack model and activities}
	\label{fig:labhack}
\end{figure}

The LabHackathon, or LabHack, model was devised by Louise Bezuidenhout and Helena Webb. It builds on hackathons and, in particular, the `ethical hackathon' model developed by members of the Human Centred Computing Theme at the University of Oxford as a practical RRI exercise~\cite{hccrri}. The overview of the LabHack model can be see in Figure~\ref{fig:labhack}. LabHacks are competitive and educational events where teams of university students compete around design challenges to build low-cost and reproducible laboratory equipment. Teams must be interdisciplinary and are required to build the kinds of basic equipment routinely used in laboratories. They are also asked to submit a range of documents, including design plans, budget and any software code. These are judged alongside their completed equipment prototypes by a panel of experts from academia industry. Assessment criteria relate to factors including quality of design, documentation, and frugalness.

During the LabHack event, participants have the opportunity to attend a variety of interactive, educational sessions as well as social events to enable networking. Following the event, teams are invited to 
share their documentation via a central event website; this enables others to use the designs to build their own versions of the equipment. 
The LabHack therefore aims to foster both immediate and long term impact. It serves as an educational and motivational opportunity for participating students, who are encouraged to learn about the design and building of laboratory equipment, work collaboratively and apply creative thinking to their challenge. 
It also seeks to empower students and educators to address  equipment shortages in learning environments, increase awareness of the Open Hardware movement and facilitate the cooperative sharing of designs and ideas. 
We envisage the LabHack as an ongoing and self-sustaining process to be applied whenever there is some equipment need, hence its representation in Figure~\ref{fig:labhack} as a cycle.


\section{LabHack Pilot in Zimbabwe}
Zimbabwe was chosen as the location for a pilot LabHack event because it experiences many of the kinds of equipment shortages and associated problems the model is designed to address. A venue and date was arranged, namely the Harare Institute of Technology, June 8-10th 2018.

\subsection{Problem definition }
We wanted to challenge participants to produce pieces of equipment that: i) are frequently used in learning laboratories for STEM; ii) have a software and hardware component; iii) have the potential to be built frugally; and iv) represent an achievable challenge for university students to research, design and build. As a result of these considerations we specified four challenges:
\begin{itemize}
	\item Challenge 1: magnetic stirrer 
    \item Challenge 2: PCR (polymerase chain reaction) machine 
    \item Challenge 3: bench top centrifuge 
    \item Challenge 4: open (if teams had a particular interest in doing so, they could design another type of equipment similar to 1-3 in terms of use and difficulty)
\end{itemize}

We also defined our judging criteria as: quality of build; quality of accompanying documentation; reproducibility of design, and frugalness of build. 

\subsection{Team formation and selection of challenge}
 We set up a website (http://labhackathon.com) with information about the event, requirements for participation and downloadable application forms. We publicised it in universities and technical clubs in Zimbabwe with assistance from our project partners, the National Biotechnology Authority of Zimbabwe and the Southern Africa Network for Biosciences, plus other local organisations.  Teams had to include at least 4 of the following disciplines: computer science, mechanical/electrical engineering, life/natural sciences, economics, marketing and other social sciences. They were encouraged to select the challenge that best fitted their expertise and interests. Application documents were reviewed and, when completed satisfactorily, teams were registered to take part. Although we initially planned the event for undergraduates, we had a number of registrations from interested young hobbyists who were not at university. In total 13 teams registered, with an average of 6 members per team.

\subsection{Prototype creation}
A particular novelty of the LabHack model is that prototype creation begins before the competition event. In a typical hackathon, little preparation is required however, the LabHack requires teams to do much of the work to design and build their equipment before the event itself. This offers many advantages. Firstly, it allows more time for teams to produce quality work as they can iterate their designs. The longer creation phase also offers an extended learning 
period during which participants can develop the skills necessary for building their prototypes. Furthermore, it means that design documents can be submitted in advance of the event providing judges with more time to assess the teams. Finally, less time is required for prototype creation during the event meaning that much of the schedule can be dedicated to learning sessions. To assist with prototype creation, each team was provided with an Arduino kit. We also encouraged participants to visit our website, where we provided links to various support sources on Open Hardware, Raspberry Pi/Arduino projects and more. 

The event itself took place over a weekend (Friday evening to Sunday afternoon) and of the 13 teams who registered, 11 attended and submitted designs. We also opened the event to observers from local schools and universities. Observers did not compete in challenges but were able to view the prototypes and take part in the interactive sessions and social events. Over 150 people attended in total. 

The event opened on the Friday afternoon with time for team set up and a welcome social event. On Saturday morning presentations were given about Open Science and Open Hardware. The teams then displayed their prototypes --- first by giving a brief spoken presentation and then in a demonstration session. The day closed with interactive workshops run by local companies and event collaborators. These included sessions on 3D printing and scanning, and using Arduinos, plus a hands-on session on building a \$10 microscope. In the evening was another social event. Sunday began with further educational sessions. In the afternoon there was another demonstration session during which the teams -- many of whom had worked on their builds overnight -- showed their finalised prototypes. A panel of judges then awarded prizes. Two main prizes of \$100 equipment vouchers were awarded for Best Prototype and Best Documentation. Smaller prizes were also given and included Most Frugal Design and Best Integration of Concepts. 

\subsection{Open Source prototypes}
Following the event, some of the winning design documents have been displayed on our project website. They have also been exhibited, alongside information about the LabHack model, at dissemination events including the `Orbit: Building in the Good conference' held at Microsoft Research Cambridge in September 2018 and the IF Oxford Science and Ideas Festival, October 2018.

\section{Lessons learned and recommendations}
We were very pleased with the outcomes of our pilot LabHack event. The teams produced excellent work working prototypes, often for less than \$100 when commercial versions would typically cost thousands of dollars. Our participants were highly motivated and enthusiastic, and we also we received very positive feedback -- as seen in Sidebar~\ref{bar:sidebar3}. At the same time we learned some important lessons that we believe are of value to the CHI community.

\textbf{Event timing}: We timed our event for the first weekend after university exams. Whilst this meant student participants were available,  
many of them did not have as much time as they wanted to prepare their prototypes in advance of the event. 

\textbf{Event organisation}: The process of planning a LabHack in Zimbabwe whilst in the UK led to a number of complications around the transfer of funds and communication with local organisations. Similar issues 
\begin{sidebar}
	
	``Thank you so much for giving us an opportunity to learn and above all being in the midst of people who appreciate that as youth we can contribute in the science community. [Our team] is not a result of a challenge you gave us but we are a result of the out cry of young innovative children in Zimbabwe and Africa as a whole. We are the future generation and we also want to take part in shaping that future. I have no words to fully describe the impact of your visit to us and hosting us in the labhackathon.'' 
	
	\vspace{1em}
	
	``Me and my team wanted to thank you guys for organising this event we had a lot of fun and we learnt so much. We will continue as team and hopefully we can do more and achieve more. As a way to thank you guys for making us realize ourselves we will work hard and take ourselves to the next level will keep you posted on our achievements.
	And again thank you guys its only because of you we became [our team].'' 
	
	\vspace{1em}
	
	``Thank you so much for appreciating the ... work we are doing. It means a lot to us and helps keep our spirits high.''
	
	\vspace{1em}
	
	``I kindly appreciate your efforts to help us realize our potential. thank you.'' 
	
	\vspace{1em}
	
	``It was great having you in Zimbabwe, we hope to see you for the next season of labhack Zim. A big thank you for arranging such an event with radical and progressive people in science.''
	
	\caption{Feedback from LabHack participants}
	\label{bar:sidebar3}
\end{sidebar}
are likely at future events held elsewhere as the bureaucratic norms of UK universities and funding agencies do not always map well onto the practices of African institutions.  We received excellent support from our local project partners and venue and in future can consolidate this by building in organisational roles for local staff members to benefit from their on-the-ground knowledge. 

\textbf{The need to be flexible}: Some challenges could not be overcome and we had to work around them. Our event was originally planned for March 2018 but the uncertainty in Zimbabwe following the political events of November 2017 meant we had to postpone. We were unexpectedly charged customs duties for the materials we brought into Zimbabwe and had to explain to participants that our planned \$10 microscope ultimately cost a little more. We were also unable to source in local Harare stores much of the hardware we planned to use in the interactive sessions. This provided an excellent lesson in the kinds of resource scarcity our participants routinely encounter and, like them, we had to think creatively - by sourcing alternative materials and adapting the sessions -  to overcome this. 

\textbf{Open Hardware}: Awareness of the Open Science and Open Hardware movements amongst our participants was low and we could do more to make relevant resources accessible to them - in particular to help them in the prototype creation phase. Additionally, despite the excellence of the prototypes and documents produced, we need to find mechanisms to support the ongoing efforts of teams to calibrate their equipment correctly and bring their plans up to a standard that means they can be shared to enable others to reproduce the equipment.

\section{Conclusions and future work}
The LabHack model draws on Open Hardware and RRI approaches to address resource scarcity in Africa by challenging participants to design and build frugal and reproducible laboratory equipment. Our pilot event demonstrated that it is a viable initiative. A second LabHack was held by our partners the Southern Africa Network for Biosciences in South Africa in November 2018 and we continue to seek funding to run further events. Our long term task is to create a sustainable network so that events can be run in multiple countries, and a central repository set up to share and store planning materials and the documentation produced by participants. We are currently discussing steps towards achieving this sustainability with funding agencies as well as regional and national collaborators.

\newpage
\begin{acks}
  We thank all the participants in LabHack Zimbabwe 2018, and special thanks to  
  \begin{marginfigure}
  	\vspace{2em}
  	\includegraphics[width=\marginparwidth]{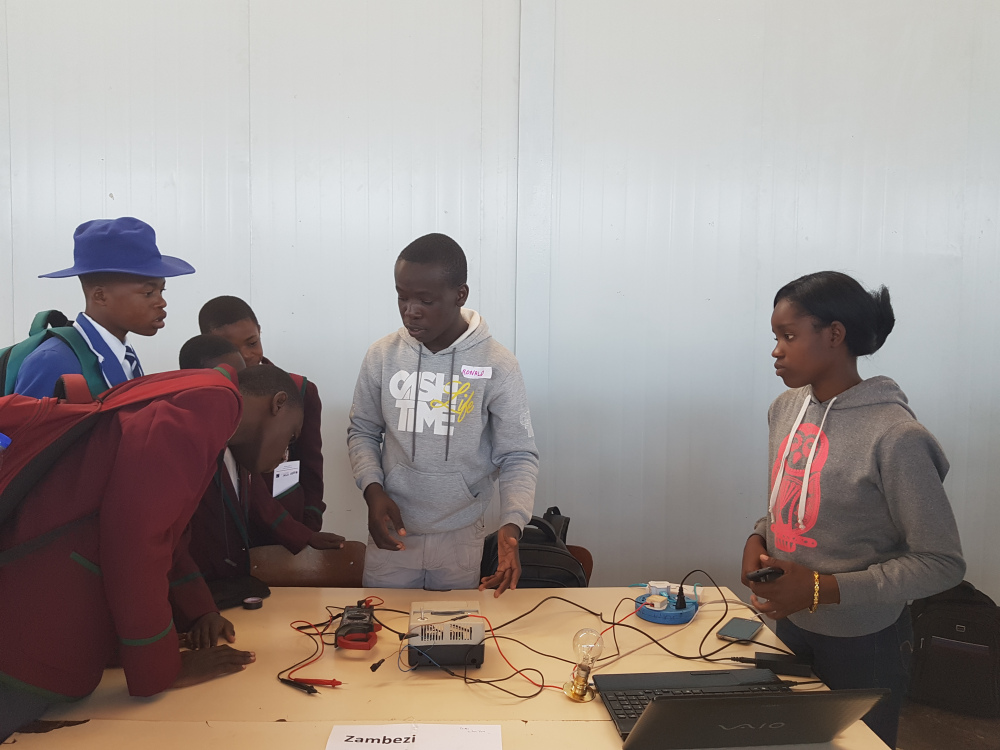}
  	\caption{A team presenting their prototype}
  	\label{fig:group}
  \end{marginfigure}
  \begin{marginfigure}
  	\vspace{2em}
  	\includegraphics[width=\marginparwidth]{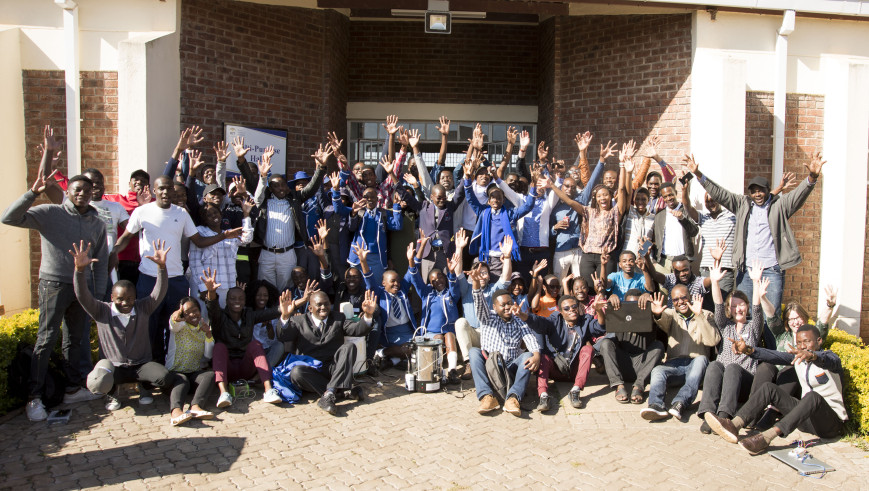}
  	\caption{LabHack Zimbabwe 2018 group photo. Photo credit to Jeffrey Barbee/Alliance Earth}
  	\label{fig:group2}
  \end{marginfigure}
  Harare Institute of Technology  (HIT), NEPAD Southern African Network of Biosciences (SANBio) and the National 
  Biotechnology Authority (NBA) for their support. Also, we 
  thank the UK EPSRC Global 
  Challenge Research Fund Institutional Sponsorship for the funding 
  provided to conduct this project.
\end{acks}

\bibliography{bibliography}
\bibliographystyle{ACM-Reference-Format}

\end{document}